\documentclass[twocolumn,prb,aps,showpacs,superscriptaddress]{revtex4}
\usepackage{graphicx}
\usepackage{dcolumn}
\usepackage{bm}
\usepackage{amsmath}
\usepackage{amsfonts}

\begin{document}

\title{Theory of nonlinear optical spectroscopy of electron spin coherence in quantum dots}
\author{Ren-Bao Liu}
\affiliation{Department of Physics, The Chinese University of Hong Kong, Shatin, N.T., Hong Kong, China}
\author{S. E. Economou}
\affiliation{Department of Physics, University of California San Diego, La Jolla, California 92093-0319}
\author{L. J. Sham}
\affiliation{Department of Physics, University of California San Diego, La Jolla, California 92093-0319}
\author{D. G. Steel}
\affiliation{The H. M. Randall Laboratory of Physics, University of Michigan, Ann Arbor, MI 48109}
\date{\today}

\begin{abstract}
We study in theory the generation and detection of electron spin coherence in
nonlinear optical spectroscopy of semiconductor quantum dots doped with single electrons.
In third-order differential transmission spectra, the inverse width of the ultra-narrow peak
at degenerate pump and probe frequencies gives the spin relaxation time ($T_1$),
and that of the Stoke and anti-Stoke spin resonances gives the effective spin dephasing time
due to the inhomogeneous broadening ($T_2^*$). The spin dephasing time excluding the
inhomogeneous broadening effect ($T_2$) is measured by the inverse width of ultra-narrow hole-burning
resonances in fifth-order differential transmission spectra.
\end{abstract}

\pacs{76.70.Hb,42.65.An, 78.67.Hc}
\maketitle

\section{Introduction}
Electron spin coherence in semiconductor quantum dots (QDs) is a
quantum effect to be exploited in emerging technologies such as
spin-based electronics (spintronics) and quantum
computation.\cite{SpintronicsQC} The electron spin decoherence is
a key issue for practical application of the electron spin
freedom and is also of fundamental interest in mesoscopic physics
and in quantum physics. The electron spin decoherence in QDs,
however, is yet poorly characterized. By convention, the spin
decoherence is classified into the longitudinal and the
transverse parts, which correspond to the spin population flip
and the Zeeman energy fluctuation processes and are usually
characterized by the relaxation time $T_1$ and the dephasing time
$T_2$, respectively. Most current experiments are carried out on
ensembles of spins, composed of either many similar
QDs\cite{Gupta_spin_Nano,Gurudev,Marie_NucleiQD,Greilich_lock} or many
repetitions of (approximately) identical measurements on a single
QD.\cite{Fujisawa_T1,Readout_spin_Kouwenhoven,Marcus_T1,Finley_spin_memory,
Gammon_opticalPump,Imamoglu_cooling,Marcus_T2,Kouwenhoven_singlet_triplet,Koppens_T2}
The ensemble measurements are subjected to the inhomogeneous
broadening of the Zeeman energy which results from the
fluctuation of the QD size, shape and compound composition (and
in turn the electron $g$-factor) and from the random distribution
of the local Overhauser field (due to the hyperfine interaction
with nuclear spins in thermal states). The inhomogeneous
broadening leads to an effective dephasing time
$T_2^*$.\cite{Merkulov_decoherence_nuclei,Semenov_Nuclear,Loss_decoherence_nuclei}
The three timescales characterizing the electron spin decoherence
can differ by orders of magnitude usually in the order $T_1\gg
T_2\gg T_2^*$. For example, in a typical GaAs QD at a low
temperature ($\lesssim 4$~Kelvin) and under a moderate to strong
magnetic field ($0.1\sim 10$~Tesla), the relaxation time $T_1$
can be in the order of
milliseconds,\cite{Fujisawa_T1,Readout_spin_Kouwenhoven,Marcus_T1,
Finley_spin_memory,Gammon_opticalPump,Imamoglu_cooling} the
dephasing time $T_2$ is up to several
microseconds,\cite{Marcus_T2,Koppens_T2,Greilich_lock} and the
effective dephasing time $T_2^*$ can be as short as a few
nanoseconds.\cite{Gurudev,Greilich_lock,Marcus_T2,Kouwenhoven_singlet_triplet}

The issue is how to measure the characteristic times of electron
spin decoherence in QDs. There have been many experiments both in
optics\cite{Marie_NucleiQD,Finley_spin_memory,Gammon_opticalPump,Imamoglu_cooling}
and in transport\cite{Fujisawa_T1,Readout_spin_Kouwenhoven,Marcus_T1}
which establish the spin relaxation time $T_1$ in QDs of
different materials. The effective dephasing time  $T_2^*$ is
also measured for QD ensembles,\cite{Gupta_spin_Nano,Gurudev,Greilich_lock,Kouwenhoven_singlet_triplet,Marcus_T2}
giving a lower bound of $T_2$. Spin echo in microwave ESR
experiments is a conventional approach to measuring the spin
decoherence time $T_2$ excluding the inhomogeneous
broadening,\cite{ESR_silicon_T2,Abe_Si,Abe_Si29} which, however,
is less feasible for III-V compound quantum dots due to the
ultrafast timescales in such systems ($T_2\lesssim 10^{-6}$~sec
and $T_2^*\lesssim 10^{-9}$~sec). Indeed, the remarkable spin
echo experiments in coupled QDs done by the Marcus group
are performed with rather long DC voltage pulses instead of
instantaneous microwave pulses.\cite{Marcus_T2} Alternatively,
picosecond optical pulses may be used to manipulate electron
spins via Raman processes\cite{Chen_Raman} and realize the spin
echo, which, however, still need to overcome the
difficulty of stabilizing and synchronizing picosecond pulses in
microsecond time-spans. A recent experiment by Greilich et al
also shows that the inhomogeneous broadening effect can be filtered
out from the spin coherence mode-locked by
a periodic train of laser pulses.\cite{Greilich_lock}

In this paper, we will study the frequency-domain nonlinear optical spectroscopy as another approach to
measuring the electron spin decoherence times. Particularly, the spin dephasing rate $T_2^{-1}$ is
correlated to the width of ultra-narrow hole-burning peaks in fifth order differential transmission (DT) spectra.
This hole-burning measurement of the spin dephasing time is analogous to the exploration of slow relaxation of
optical coherence in atomic systems by the third-order hole-burning spectroscopy.\cite{Steel_1987} Here the fifth order
nonlinearity is needed because the creation of spin coherence by Raman processes involves at least two orders of
optical field and hole-burning two more. The state-of-the-art spectroscopy already has the ultra-high resolution
(much better than MHz-resolution) to resolve the slow spin decoherence in microsecond or even millisecond
timescales.\cite{Steel_1985,Strickland_2000,Ohno_2004}

The organization of this paper is as follow:
After this introductory section, Sec.~\ref{theory} describes the model for QD system and the master-equation
approach to calculating the nonlinear optical susceptibility. Sec.~\ref{results} presents the results and discussions.
Sec.~\ref{conclusion} concludes this paper. The solution of the master equation in frequency domain is presented in
Appendix~\ref{Append_master}.

\section{Model and theory}
\label{theory}

\begin{figure}[b]
\begin{center}
\includegraphics[width=6.2cm, height=3.5cm, bb=55 460 540 735,clip=true]{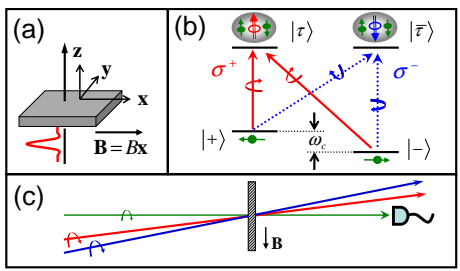}
\end{center}
\caption{(Color online). Schematics of (a) the quantum dot, (b) the selection rules for optical transitions, and
(c) the optical detection geometry.}
\label{Fig_Schematics}
\end{figure}

The system to be studied is a semiconductor QD doped with a single electron.
The geometry of the QD under an external magnetic field and optical excitation is shown in Fig.~\ref{Fig_Schematics}
(a) and (c). The QD is assumed of a shape with small thickness in the growth direction and relatively
large radius in the lateral directions, as in the usual cases of fluctuation QDs
and self-assembled QDs.\cite{Gurudev,Marie_NucleiQD,Finley_spin_memory,Gammon_opticalPump,Imamoglu_cooling}
To enable the generation and manipulation of the electron spin coherence
through Raman processes, a magnetic field is applied along a lateral direction ($x$-axis).
The propagation directions of the pump and probe laser beams are close to the growth direction ($z$-axis).
The two electron spin states $|\pm\rangle$ are split by the magnetic field with Zeeman energy $\omega_0$.
The strong confinement along the $z$-axis induces a large splitting between the heavy hole and the light hole states,
thus the relevant exciton states are the ground trion states $|\tau\rangle$ and $|\bar{\tau}\rangle$ which consist
of two electrons (including the doped one and one created by optical excitation) in the singlet spin state and
one heavy hole in the spin state $|+3/2\rangle$ and $|-3/2\rangle$ (quantized along the $z$-axis with nearly
zero Zeeman splitting), respectively. Similarly, we can also neglect the excitation of higher lying trions,
bi-exciton and multi-exciton states since the energy of adding an exciton in each case is
well separated from energy of the lowest trion states.
The selection rules for the optical transitions are determined by the (approximate) conservation of the
angular momentum along the growth direction so that a circularly polarized light with polarization
$\sigma^{+}$ or $\sigma^-$ connects the two electron spin states to the trion state $|\tau\rangle$
or $\bar{\tau}\rangle$, respectively [see Fig.~\ref{Fig_Schematics} (b)].
The relaxation processes in the system are parameterized by the exciton recombination rate $\Gamma_1$, the
exciton dephasing rate $\Gamma_2$, the spin relaxation rate $T_1^{-1}$, and the spin dephasing
rate $T_2^{-1}$. The inhomogeneous broadening leads to a random component $\epsilon$ to the Zeeman splitting:
$\omega_c=\omega_0+\epsilon$, which is assumed of Gaussian distribution
$g(\epsilon)=e^{-(\epsilon T_2^*)^2/2}/(\sqrt{2\pi}/T_2^*)$.
The hole spin relaxation is neglected since it is extremely slow when the hole
is confined in the trion states.\cite{Imamoglu_cooling} The theory presented here can be
extended straightforwardly to include the hole spin relaxation, the light-hole states,
the hole mixing effect (which leads to the imperfection in the selection rules), the
multi-exciton states, the inhomogeneous broadening of the
trion states, and so on, but we expect no qualitative modification of the resonance features related
to the electron spin coherence in the nonlinear optical spectra. For the interests of simplicity, we
shall consider only $\sigma^+$-polarized optical field (extension to other polarization configurations
may provide some flexibility for experiments and is trivial in the theoretical part).
Thus the model is reduced to a $\Lambda$-type three-level system consisting of the two
electron spin states $|\pm\rangle$ and the trion state $|\tau\rangle$.
The $\Lambda$-type three-level model, in spite of its simplicity, is the basis of a wealth
of physical effects including electromagnetically induced transparency,\cite{EIT}
lasing without inversion,\cite{LaserNoInv} and stimulated Raman adiabatic passage,\cite{Bergmann_1998}
and has been successfully applied to study transient optical signals of doped quantum
dots.\cite{Gurudev,Economou_2005}

The dynamics of the system is described in the density matrix formalism with $\rho_{\alpha,\beta}$ being the
density matrix elements between the states $|\alpha\rangle$ and $|\beta\rangle$. The optical excitation
and the relaxation are accounted for in the master equation as
\begin{subequations}
\begin{eqnarray}
\hskip -1cm \partial_t \rho_{\tau,\pm} &=& -i({\mathcal
E}_g\mp\omega_{c}/2-i\Gamma_{2})\rho_{\tau,\pm} \nonumber \\
  && -i E(t)\rho_{\tau,\tau}  +i E(t)\rho_{\mp,\pm}    +i E(t)\rho_{\pm,\pm}   ,\\
 \hskip -1cm \partial_t \rho_{\tau,\tau} &=&
-2\Gamma_{1}\rho_{\tau,\tau}
   +2\Im\left[ E^*(t)\rho_{\tau,+}+E^*(t)\rho_{\tau,-}\right],\\
\hskip -1cm \partial_t  \rho_{\pm,\pm} &=&
-\left(p_{\mp}\rho_{\pm,\pm}-p_{\pm}\rho_{\mp,\mp}\right)/T_1
+\Gamma_{1}\rho_{\tau,\tau} \nonumber \\
     &&  -{2}\Im\left[ E^*(t)\rho_{\tau,\pm}\right],\\
\hskip -1cm  \partial_t \rho_{\pm,\mp}&=&
\Gamma_{1}\rho_{\tau,\tau}-i(\pm
\omega_{c}-i/T_{2})\rho_{\pm,\mp} \nonumber \\
     &&  +i E^*(t)\rho_{\tau,\mp}    -i
    E(t)\rho^*_{\tau,\pm},\label{SGC}
\end{eqnarray}
\label{master}
\end{subequations}
where ${\mathcal E}_g$ is the energy gap,
and $p_{\pm}$ is the equilibrium population of the spin states
in absence of the optical excitation, and the optical field $E(t)=\sum_jE_je^{-i\Omega_j t}$
contains different frequency components.
The transition dipole moment is understood to be absorbed into the field quantities.
In the rotating wave reference frame, the energy gap ${\mathcal E}_g$ is set to be zero and
the optical frequencies $\Omega_j$ are measured from the gap.
The first term in the righthand side of Eq.~(\ref{SGC}) is the spin coherence generated by
spontaneous emission,\cite{Economou_2005,Java_SGC,read_Gammon}
which has been demonstrated in time-domain experiments with significant effects on spin beats.\cite{Gurudev}
We will show that it produces extra resonances in fifth-order DT spectra.

To calculate the nonlinear optical
susceptibility, the master equation is obtained in the frequency-domain (as given in Appendix~\ref{Append_master}).
With the spectrum of the optical field given by $E(\Omega)=\sum_j 2\pi E_j\delta\left(\Omega-\Omega_j\right)$,
the density matrix can be expanded as
\begin{eqnarray}
\rho_{\alpha,\beta}(\Omega) &= & \sum_{j,\ldots,k;m,\ldots,l}2\pi  E_j\cdots E_k E_m^*\cdots E_l^*  \nonumber \\
 && \times \rho_{\alpha,\beta}^{(j\cdots k \bar{m}\cdots \bar{l})} \delta\left(\Omega-
\Omega_{j\cdots k \bar{m}\cdots \bar{l}}\right),
\end{eqnarray}
where $\Omega_{j\cdots k \bar{m}\cdots \bar{l}}\equiv \Omega_j+\cdots+\Omega_k-(\Omega_m+\cdots +\Omega_l)$.
The derivation of the density matrix component $\rho^{(j\cdots k \bar{m}\cdots \bar{l})}$
up to the fifth order is lengthy but straightforward.
The final result is averaged with the inhomogeneous broadening distribution $g(\epsilon)$.

\section{Results and discussions}
\label{results}

The linear optical susceptibility is given by
\begin{eqnarray}
\rho^{(j)} =\int \frac{g(\epsilon)}{\Omega_j-{\mathcal E}_g\pm\omega_c/2+i\Gamma_2}d\epsilon.
\end{eqnarray}
In fluctuation GaAs QDs, the exciton dephasing is much faster than the effective spin dephasing
due to the inhomogeneous broadening ($\Gamma_2^{-1}\lesssim 0.1$~ns $\ll T_2^*\sim 10$~ns),\cite{Gurudev}
so the resonance width in linear optical spectra is usually dominated by the trion state
broadening, revealing little information about the spin decoherence.

\begin{figure}[b]
\begin{center}
\includegraphics[width=5.1cm, height=2.25cm, bb=100 615 440 765,clip=true]{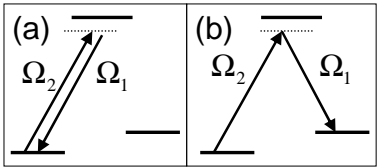}
\end{center}
\caption{ Schematics of Raman processes generating (a) the spin population and (b) the off-diagonal
spin coherence.}
\label{Fig_Raman}
\end{figure}

\begin{figure}[t]
\begin{center}
\includegraphics[width=7.7cm, height=5.1cm, bb=95 463 535 753,clip=true]{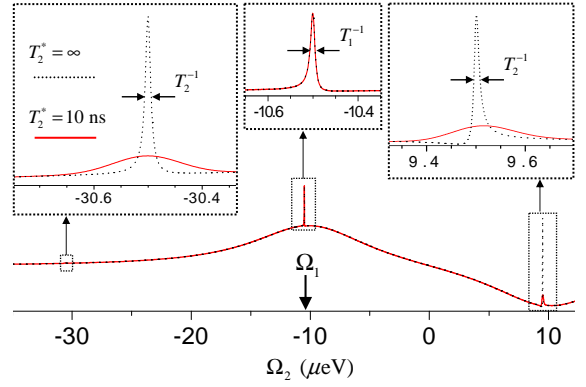}
\end{center}
\caption{(Color online). Third-order DT spectra of QDs doped with single electrons.
The parameters are chosen such that the Zeeman energy $\omega_0=20$~$\mu$eV, the spin population $p_{\pm}=0.5$,
the pump frequency $\Omega_1={\mathcal E}_g-10.5$~$\mu$eV, $\Gamma_1=5$~$\mu$eV ($\Gamma_1^{-1}\approx 0.12$~ns),
$\Gamma_2=6$~$\mu$eV ($\Gamma_2^{-1}\approx 0.1$~ns), $T_1=100$~ns, $T_2=100$~ns, and $T_2^*=10$~ns or artificially
set to be $\infty$ for the solid and dotted lines, respectively.
The insets are enlarged plots showing details of the resonances.}
\label{Fig_chi3}
\end{figure}

In third-order optical response, the population and off-diagonal coherence of the electron spin are
generated by the Raman processes
\begin{subequations}
\begin{eqnarray}
&& \hskip -1cm \rho_{\pm,\pm}\stackrel{E_2}{\longrightarrow}
\rho^{(2)}_{\tau,\pm} \stackrel{E_1^*}{\longrightarrow}
\rho^{(2\bar{1})}_{\pm,\pm}\propto \left(\Omega_{2\bar{1}}+i/T_1\right)^{-1},
\label{population}\\
&& \hskip -1cm \rho_{\pm,\pm}\stackrel{E_2}{\longrightarrow}
\rho^{(2)}_{\tau,\pm} \stackrel{E_1^*}{\longrightarrow}
\rho^{(2\bar{1})}_{\mp,\pm}\propto \left(\Omega_{2\bar{1}}\pm\omega_c+i/T_1\right)^{-1},
\label{offdiagonal}
\end{eqnarray}
\label{spincoherence}
\end{subequations}
corresponding to the illustrations in Fig.~\ref{Fig_Raman} (a) and (b), respectively.
Another optical field with frequency $\Omega_1$ brings the second order spin coherence into the
third order optical coherence $\rho^{(21\bar{1})}_{\tau,\pm}$. The DT spectrum as a function of
the pump frequency $\Omega_1$ and the probe frequency $\Omega_2$ is
\begin{equation}
S_{\rm DT}\left(\Omega_2,\Omega_1\right)\propto -\Im \left[\rho^{(21\bar{1})}_{\tau,+}+\rho^{(21\bar{1})}_{\tau,-}\right],
\end{equation}
which presents ultra-narrow resonances around $\Omega_{2\bar{1}}=\pm\omega_0$ and $\Omega_{2\bar{1}}=0$,
with resonance width $T_2^{-1}$ and $T_1^{-1}$, related to the spin population and off-diagonal
coherence in Eq.~(\ref{population}) and (\ref{offdiagonal}), respectively.
Such resonances are shown in Fig.~\ref{Fig_chi3}. Thus the spin dephasing time $T_2$ and
relaxation time $T_1$ are measured. But when the inhomogeneous
broadening is included, since usually $T_2\gg T_2^*$, the Stoke and anti-Stoke Raman
resonances at $\Omega_{2\bar{1}}=\pm\omega_0$ will be smeared to be a peak
resembling the inhomogeneous broadening distribution as
\begin{eqnarray}
\int \rho^{(2\bar{1})}_{\pm,\mp} g(\epsilon) d\epsilon \sim -i\pi
g\left(\Omega_{2\bar{1}}\mp\omega_0\right).
\end{eqnarray}
The effect of the inhomogeneous broadening is clearly seen in Fig.~\ref{Fig_chi3}.
So in usual cases, the third-order DT spectra measure the $T_2^*$ instead of the
$T_2$. The resonance at degenerate pump and probe frequencies ($\Omega_{2\bar{1}}=0$)
is related to the spin population and is immune to the random distribution of the electron Zeeman energy.
So the spin relaxation time $T_1$ can be deduced from the third order DT spectra,
regardless of the inhomogeneous broadening.

\begin{figure}[b]
\begin{center}
\includegraphics[width=7.4cm, height=2.4cm, bb=85 665 455 785,clip=true]{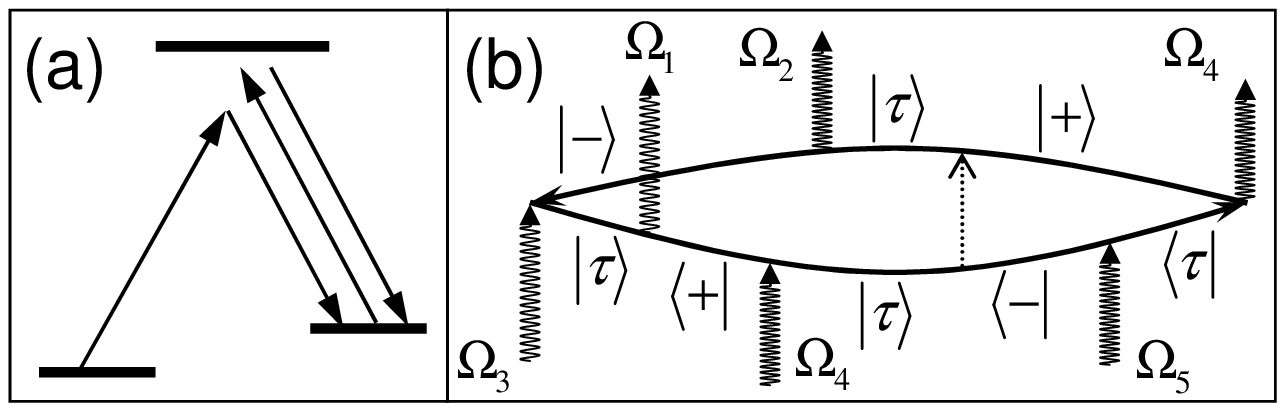}
\end{center}
\caption{(a) Schematics for a fourth-order optical process that generate spin coherence with
a double resonance structure. (b) The Feynman diagram for the fifth order optical response
involving the spontaneous emission, in which the optical field and the vacuum field
are represented by the wavy arrows and the dotted arrow, respectively.}
\label{Fig_chi5diagram}
\end{figure}

To measure the spin dephasing time excluding the inhomogeneous broadening effect,
the fifth order nonlinearity can be used. In the fifth order optical response,
the spin coherence in the fourth order of optical field has very rich
resonance structures. For instance, a double resonance like
\begin{eqnarray}
\rho^{(43\bar{2}\bar{1})}_{+,-} \sim
\left(\Omega_{3\bar{1}}-\omega_c+i/T_2\right)^{-1}
\left(\Omega_{4 3\bar{2}\bar{1}}-\omega_c+i/T_2\right)^{-1},
\label{hole_burning}
\end{eqnarray}
arises from the excitation pathway
\begin{eqnarray}
\rho_{-,-}\stackrel{E_3}{\longrightarrow}\rho^{(3)}_{\tau,-}
\stackrel{E_1^*}{\longrightarrow}\rho^{(3\bar{1})}_{+,-}
\stackrel{E_4}{\longrightarrow}\rho^{(43\bar{1})}_{\tau,-}
\stackrel{E_2^*}{\longrightarrow}\rho^{(43\bar{2}\bar{1})}_{+,-},
\label{pathway_chi5A}
\end{eqnarray}
as depicted in Fig.~\ref{Fig_chi5diagram}~(a). The double resonance will manifest itself in a
two-dimensional DT spectrum as an ultra-narrow peak at $\Omega_{3\bar{1}}=\Omega_{43\bar{2}{1}}=\omega_c$
with width $\sim T_2^{-1}$. When the inhomogeneous broadening is included, the ultra-narrow
resonance will be smeared into a broadened peak along the direction
$\Omega_{3\bar{1}}=\Omega_{43\bar{2}{1}}$ with width $\sim 1/T_2^*$. But in the perpendicular direction (defined by
$\Omega_{3\bar{1}}=-\Omega_{43\bar{2}{1}}$), the peak width remains unchanged.
So when $\Omega_{43\bar{2}{1}}$ is fixed around $\omega_0$ and $\Omega_{3\bar{1}}$ is scanned, or vice versa,
the DT spectrum will present a sharp peak whose width measures the inverse spin dephasing time $T_2^{-1}$.
This peak has the character of hole-burning: The first frequency difference acts just as a selection of
QDs with Zeeman energy $\omega_c=\Omega_{43\bar{2}\bar{1}}$ from the inhomogeneously broadened ensemble.
The hole-burning resonance resulting from the excitation pathway in Eq.~(\ref{pathway_chi5A}), however,
emerges together with the resonance associated with the spin population $(\Omega_{4\bar{2}}+i/T_1)^{-1}$
as given in Eq.~(\ref{population}).
To avoid the complication of mixing two types of resonance structures,
we would rather make use another mechanism for spin coherence generation, namely, the spontaneous emission
that connects the trion state to the two spin states through the vacuum field [related to the first term in
the righthand side of Eq.(\ref{SGC})].\cite{Gurudev,Economou_2005,Java_SGC,read_Gammon}

\begin{figure}[t]
\begin{center}
\includegraphics[width=7.7cm, height=5.2cm, bb=55 452 545 782,clip=true]{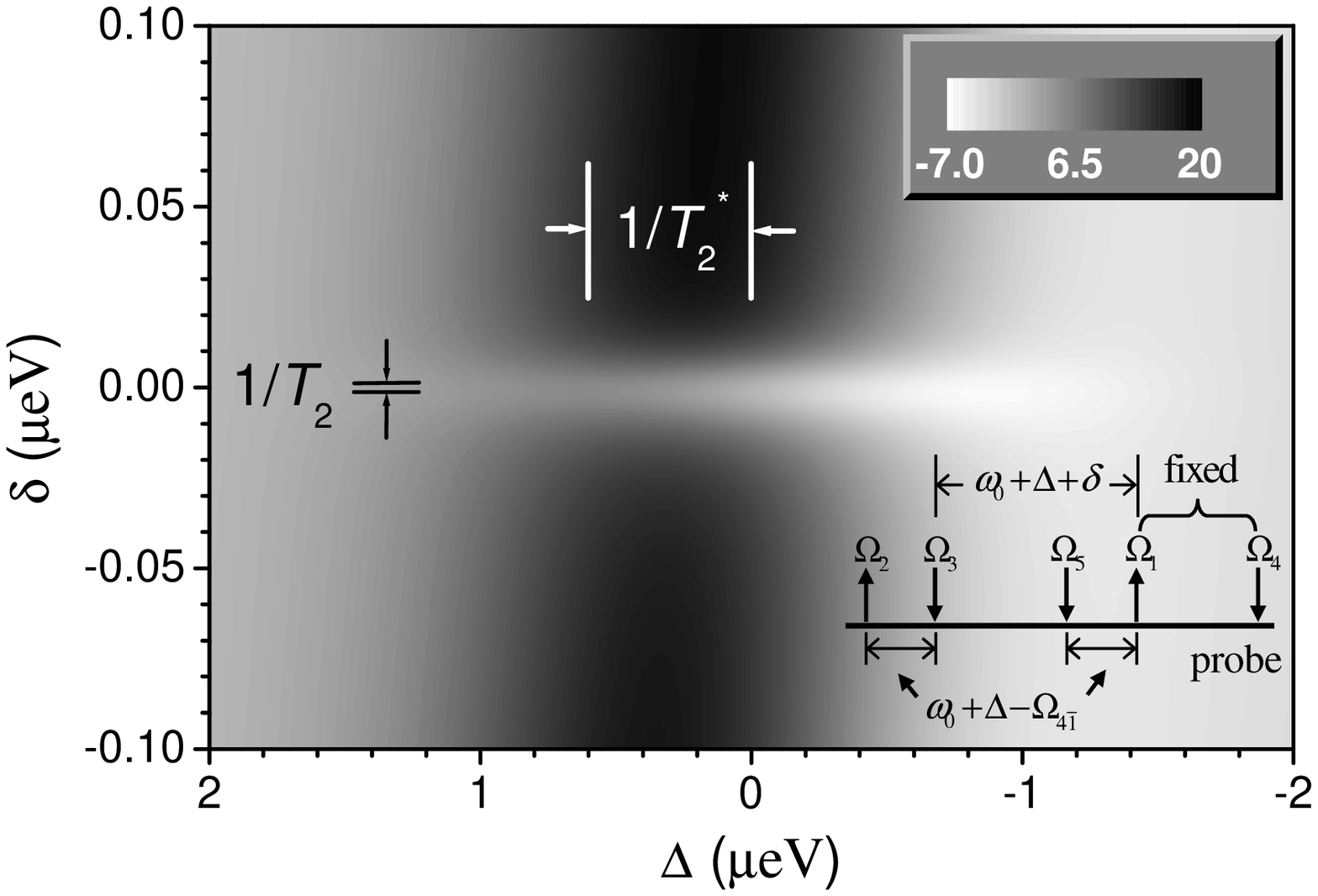}
\end{center}
\caption{Contour plot of the fifth-order DT spectrum of the QDs as a function of
$\Delta\equiv \Omega_{43\bar{2}{1}}-\omega_0$ and $\delta \equiv \Omega_{1\bar{3}}-\Omega_{43\bar{2}{1}}$.
The probe frequency is fixed to be $\Omega_4=22$~$\mu$eV, the pump frequency $\Omega_1$ is fixed
at $9$~$\mu$eV, and the other three pump frequencies are scanned with
$\Omega_{53\bar{2}\bar{1}}=0$ (which makes
$\Omega_{1\bar{5}}=\Omega_{3\bar{2}}=\omega_0+\Delta-\Omega_{4\bar{1}}$ and
$\Omega_{1\bar{3}}=\Omega_{5\bar{2}}=\omega_0+\Delta+\delta$, as indicated in the inset).
The Zeeman energy $\omega_0=20$~$\mu$eV, the spin population $p_{\pm}=0.5$, and the
relaxation rates are such that $\Gamma_1=5$~$\mu$eV ($\Gamma_1^{-1}\approx 0.12$~ns),
$\Gamma_2=6$~$\mu$eV ($\Gamma_2^{-1}\approx 0.1$~ns), $T_1=100$~ns, $T_2=100$~ns, and $T_2^*=1$~ns.}
\label{Fig_chi53D}
\end{figure}

\begin{figure}[t]
\begin{center}
\includegraphics[width=7.3cm, height=10cm, bb=55 115 545 785,clip=true]{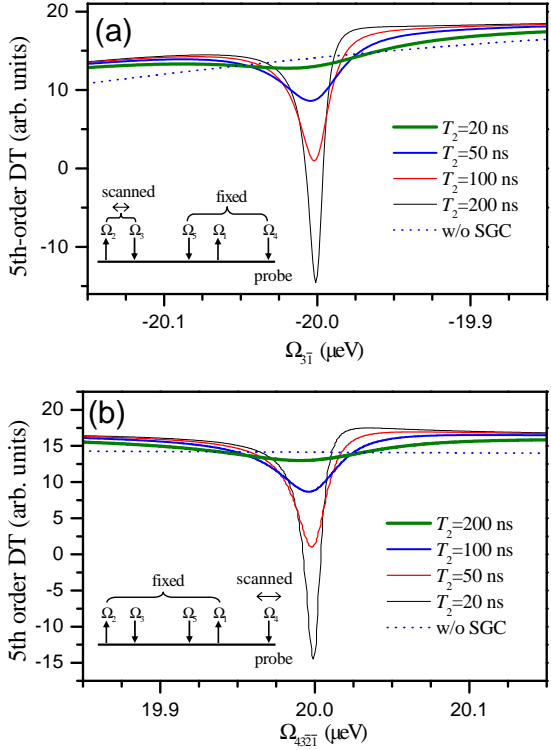}
\end{center}
\caption{(Color online) (a) The sectioned plot of Fig.~\ref{Fig_chi53D} with
$\Delta=0$ (i.e., $\Omega_{43\bar{2}\bar{1}}=\omega_0$).
(b) The fifth-order DT signal as a function of the probe frequency with pump frequencies fixed to be such that
$\Omega_1=9$~$\mu$eV, $\Omega_{5}=2$~$\mu$eV, and $\Omega_{1\bar{3}}=\Omega_{5\bar{2}}=\omega_0=20$~$\mu$eV.
In both figures, the spin dephasing time $T_2=20$, 50, 100 and 200~ns for the solid curves from top to bottom,
and the dotted line is calculated with the spontaneously generated spin coherence artificially
switched off (for $T_2=100$~ns). The parameters are the same as in Fig.~\ref{Fig_chi53D}.}
\label{Fig_chi5}
\end{figure}

The generation of spin coherence in the fifth order optical response involving the
spontaneous emission can take a quantum pathway like
\begin{equation}
\rho_{-,-}\stackrel{E_3}{\longrightarrow}\rho^{(3)}_{\tau,-}
\stackrel{E_1^*}{\longrightarrow}\rho^{(3\bar{1})}_{+,-}
\stackrel{E_4}{\longrightarrow}\rho^{(43\bar{1})}_{\tau,-}
\stackrel{E_2^*}{\longrightarrow}\rho^{(43\bar{2}\bar{1})}_{\tau,\tau}
\stackrel{\Gamma_1}{\dashrightarrow}\rho^{(43\bar{2}\bar{1})}_{-,+},
\label{pathway_SGC}
\end{equation}
where the last step is the spontaneous emission. This optical process is illustrated by
the Feynman diagram in Fig.~\ref{Fig_chi5diagram}~(b). The spin coherence generated by
the spontaneous emission and that by optical excitation can have opposite spin indices
[$\rho^{(3\bar{1})}_{+,-}\longrightarrow \rho^{(43\bar{2}\bar{1})}_{-,+}$],
which is impossible in quantum pathways without the spontaneous emission
[as can be seen from Fig.~\ref{Fig_chi5diagram} (a)].
Thus the double resonance becomes
\begin{eqnarray}
\rho^{(43\bar{2}\bar{1})}_{-,+}\sim \frac{\Gamma_1/\left(\Omega_{4 3\bar{2}\bar{1}}+i2\Gamma_1\right)}
{\left(\Omega_{3\bar{1}}+\omega_c+iT_2^{-1}\right)
\left(\Omega_{4 3\bar{2}\bar{1}}-\omega_c+iT_2^{-1}\right)},
\end{eqnarray}
which is well separated from the spin population resonance.
The spectrum is measured by fixing $\Omega_{43\bar{2}{1}}$
to be the hole-burning frequency $\omega_0+\Delta$ with $\Delta\lesssim1/T_2^*$ and
fine-tuning $\Omega_{1\bar{3}}$ to be $\Omega_{43\bar{2}{1}}+\delta$.
As shown in the inset of Fig.~\ref{Fig_chi53D},
the optical frequencies can be configured such that $\Omega_4$ and $\Omega_1$ are fixed,
$\Omega_{3}$ are red-shifted by $\omega_0+\Delta+\delta$ from $\Omega_1$, and $\Omega_5$ and
$\Omega_2$ are red-shifted by $\omega_0+\Delta-\Omega_{4\bar{1}}$ from $\Omega_1$ and $\Omega_3$, respectively.
Thus the fifth-order optical response $\rho^{(543\bar{2}\bar{1})}_{\tau,\pm}$ oscillates at the probe
frequency $\Omega_4$, which enables the signal to be measured in the DT setup instead of six-wave mixing ones.
We note that the resonance due to the spin population $\sim \left(\Omega_{53\bar{2}\bar{1}}+i/T_1\right)^{-1}$
contributes only a constant background since $\Omega_{53\bar{2}\bar{1}}\equiv 0$ in the above frequency
configuration. As shown in Fig.~\ref{Fig_chi53D} which plots the fifth order DT spectrum as a
function of $\Delta\equiv \Omega_{43\bar{2}\bar{1}}-\omega_0$
and the fine tuning $\delta\equiv \Omega_{1\bar{3}}-\Omega_{43\bar{2}\bar{1}}$,
a very narrow hole in the spectrum as a function of $\Omega_{3\bar{1}}$ is burnt around
$\Omega_{43\bar{2}\bar{1}}$, with width given by $T_2^{-1}$.
Along the direction $\Omega_{3\bar{1}}=-\Omega_{43\bar{2}\bar{1}}$ ($\delta=0$), the resonance
is extended by the inhomogeneous broadening as expected. Sectioned plots of the DT signal with fixed
$\Delta$ are shown in Fig.~\ref{Fig_chi5}~(a) for various spin dephasing time.
The resonance width is given by the spin dephasing rate, demonstrating unambiguously
that the $T_2$ is measured by the hole burning effect. The hole burning resonance can
also be detected by varying the probe frequency with the pump frequencies fixed, as
demonstrated in Fig.~\ref{Fig_chi5}~(b). The role of the spontaneous emission-generated spin coherence
is verified by the absence of the ultra-narrow resonance with artificial switch-off of
the relevant term in Eq.~(\ref{SGC}).

\section{Conclusions}
\label{conclusion}

The spin coherence can be generated and detected in nonlinear optical spectroscopy of quantum dots doped with
single electrons, which is studied in this paper up to the fifth order nonlinearity with a $\Lambda$-type
three-level model. The electron spin coherence is generated by the optical field through Raman
processes as well as by spontaneous emission of the trion. The spin population and off-diagonal coherence
demonstrate themselves in third-order differential transmission spectra as ultra-narrow resonances.
The inhomogeneous broadening smears out the sharp Stoke and anti-Stoke peaks related to the off-diagonal
spin coherence. Thus the spin relaxation time $T_1$ and the effective dephasing time $T_2^*$ are measured
by the third-order spectra. In the fifth-order optical response, the generation of the spin coherence
by both second- and fourth-order optical processes leads to double resonance structures in two-dimensional DT
spectra, which are smeared by the inhomogeneous broadening along one direction in the frequency space
but presents ultra-narrow hole-burning resonances along the perpendicular direction.
So the spin dephasing time $T_2$ is measured as the inverse width of the hole burning peak.
The spontaneous emission-generated spin coherence\cite{Gurudev,Economou_2005} is useful to
produce hole burning resonances well separated from the spin-population resonances in the fifth-order spectra.
The frequencies of the optical field can be configured properly to enable the detection of the signal in the DT
setup instead of the multi-wave mixing ones. In practice, the pump and probe frequencies may be generated from
a single continuous-wave laser source by, e.g., acousto-optical modulation.\cite{Ohno_2004}
Since the ultra-narrow hole burning peaks are rather insensitive to the global shift of the laser frequencies
and variation of the hole-burning frequency, non-stabilized laser sources may be used to resolve the slow
spin decoherence.\cite{Ohno_2004}

\begin{acknowledgments}
This work was partially supported by the Hong Kong RGC Direct
Grant 2060284 and by ARO/NSA-LPS.
\end{acknowledgments}

\begin{appendix}

\section{Solution of the master equation}
\label{Append_master}
The master equation in Eq.~(\ref{master}) can be solved in the frequency domain by Fourier transformation to be
\begin{widetext}
\begin{subequations}
\begin{eqnarray}
&& {\rho}_{\tau,\pm}(\Omega) =\int \frac{
    - E(\Omega-\omega){\rho}_{\pm,\pm}(\omega)
    - E(\Omega-\omega){\rho}_{\mp,\pm}(\omega)
    + E(\Omega-\omega){\rho}_{\tau,\tau}(\omega)}
{\Omega-{\mathcal E}_g {\pm}\omega_c/2+i\Gamma_{2}}\frac{d\omega}{2\pi},\\
&&
{\rho}_{\tau,\tau}(\omega)=\sum_{\pm}\int\frac{
     + E^*(\Omega-\omega){\rho}_{\tau,\pm}(\Omega)
     - E(\Omega+\omega){\rho}^*_{\tau,\pm}(\Omega)}
     {\omega+i2\Gamma_{1}}
     \frac{d\Omega}{2\pi},
 \\
&&{\rho}_{\pm,\pm}(\omega)={p_{\pm}}2\pi\delta(\omega)
   -\left(\omega+i\Gamma_1+ip_{\pm}/T_1\right)\int\frac{
    E^*(\Omega-\omega){\rho}_{\tau,\pm}(\Omega)
   -E(\Omega+\omega){\rho}^*_{\tau,\pm}(\Omega)
   }
   {\left(\omega+i/T_{1}\right)\left(\omega+i2\Gamma_1\right)}
   \frac{d\Omega}{2\pi}
\nonumber \\ && \phantom{{\rho}_{\pm,\pm}(\omega)={p_{\pm}}2\pi\delta(\omega)}
  +\left(i\Gamma_1-ip_{\pm}/T_1\right) \int\frac{
    E^*(\Omega-\omega){\rho}_{\tau,\mp}(\Omega)
   -E(\Omega+\omega){\rho}^*_{\tau,\mp}(\Omega)
   }
   {\left(\omega+i/T_{1}\right)\left(\omega+i2\Gamma_1\right)}
   \frac{d\Omega}{2\pi}
  , \\
&& {\rho}_{+,-}(\omega)=\frac{i{\Gamma_{1}}{\rho}_{\tau,\tau}(\omega)}
{\omega-\omega_{c}+i/T_{2}}
   + \int\frac{- E^*(\Omega-\omega){\rho}_{\tau,-}(\Omega)
    + E(\Omega+\omega){\rho}^*_{\tau,+}(\Omega)}{\omega-\omega_{c}+i/T_{2}}
    \frac{d\Omega}{2\pi}.
\end{eqnarray}
\end{subequations}
\end{widetext}
\end{appendix}

%\bibliography{references}

\end{document}